\documentstyle{article}

\def\PL{{\it Phys. Lett.\ }}

\def\be{\begin{equation}}
\def\eq{\end{equation}}

\def\q2{{\rm QCD}_{2}}
\def\q4{{\rm QCD}_{4}}

\begin{document}
\vspace{20mm}
\begin{center}
{\LARGE  Light-Cone Quantization and Renormalization 
 of } \\
{\LARGE Large-$N$ Scalar Matrix Models} \\ 
\vspace{20mm}
{\bf F. Antonuccio} \\
\vspace{10mm}
{\em Max-Planck-Institut f\"{u}r Kernphysik, 69029 Heidelberg, Germany
\\ and \\
Institute for Theoretical Physics, University of Heidelberg,
69120 Heidelberg, Germany}
\end{center}
\vspace{20mm}
\begin{abstract}
We discuss the issue of renormalization and
the derivation of effective interactions for light-cone
Hamiltonians in the context of large-$N$ 
scalar matrix models  
with $\Phi^3$ interactions. 
For various space-time dimensions $D \geq 3$, 
we deduce appropriate mass, coupling constant, and wavefunction
renormalizations which are necessary for finiteness of
the Hamiltonian at leading order.
We also outline how higher order corrections may be derived
within this framework, and discuss the relevance 
of this approach in the light-cone quantization of gauge theories.  
\end{abstract}
\newpage

\baselineskip .25in

\section{Introduction}
The important role of light-cone quantization as a 
possible non-perturbative approach towards
solving quantum field theories has been known for some
time. Traditionally, attention has focused  on
Quantum Chromodynamics (QCD), and more recently, on QCD 
inspired models.
Further details, 
and a more extensive list of references 
than the one provided in this note, can be found 
in various existing review articles \cite{rev1,rev2,rev3}.
In very recent times, a number of remarkable conjectures 
on a  possible non-perturbative formulation
of M theory involving light-cone dynamics have been made 
\cite{m1,m2,m3}.
In these developments, it is apparent 
that a correct
understanding of matrix field theories 
formulated on the light-cone might be helpful
in shedding further light on this subject.

Unquestionably, light-cone quantization
has been very successful in handling $1+1$ dimensional
field theories, and complete numerical solutions may
be obtained in many cases from discretizing -- and subsequently
diagonalizing -- the light-cone Hamiltonian. This
procedure is usually referred to as 
Discretized Light-Cone Quantization 
\cite{dlcq}. In higher dimensions, the main difficulty
appears to be the renormalization of the light-cone 
Hamiltonian, although various non-perturbative schemes
have been proposed \cite{wilson,wegner,hiller}.   
Properties such as confinement and
chiral symmetry breaking in light-cone 
field theories are expected to
emerge in a {\em non-perturbative} treatment 
of the renormalization group equations. 
At the present time, a complete solution to this problem
is still lacking.

In this short note, we turn our attention to the {\em perturbative}
renormalization of
the light-cone Hamiltonian --
and the derivation of effective interactions -- by studying 
large $N$ scalar matrix models,  with $\Phi^3$ interactions,
in $D \geq 3$ space-time
dimensions. We shall see that there exists a strategy
that is natural to the light-cone coordinate frame\footnote{
The light-cone frame should be distinguished from
the Infinite Momentum Frame. See \cite{m3}.}.
We also compare 
our results with those 
given by the usual
equal-time covariant approach to renormalization.

The content of this note is divided up as follows. In Section
\ref{themodel} we formulate $\Phi^3$ matrix 
field theory on the light-cone,
and give expressions for the light-cone Hamiltonian $P^-$
and conserved total momenta $(P^+,{\bf P}^{\perp})$.
Light-cone quantization of this field theory is performed
in Section \ref{lcq}, and the associated bound state 
integral equations are presented in Section \ref{boundstate}.
What then follows is a small momentum fraction-$x$ analysis
of these equations, which enables one to study the
ultraviolet properties of the theory. Discussion
of a renormalization prescription for eliminating 
ultraviolet divergences appears in Section \ref{renorm}. 
In Section \ref{tamm}, we consider the  
impact of
our analysis in the context of numerical approaches -- in 
particular, Discretized Light-Cone Quantization -- and
assess the relevance of further analytical work in this subject.
We conclude with a summary of our results, and discuss
(near) future prospects for gauge theories.  

\section{The $\Phi^3$ Matrix Model} \label{themodel}
Consider the $D$ dimensional field theory described by the action
\begin{equation}
 S= \int d^D{\bf x} \hspace{1mm}\mbox{Tr}\left[ \frac{1}{2} \partial_{\mu} \Phi
 \partial^{\mu} \Phi - \frac{1}{2}m^2 \Phi^2 + \frac{\lambda}
{3 \sqrt{N}} \Phi^3 \right],
\label{phi3model}
\end{equation}  
where $\Phi({\bf x})$ is an $N \times N$ Hermitian matrix field on 
$D$ dimensional Minkowski space ${\bf x} \equiv (x^0,x^1,\dots,x^{D-1})$.
Note that the action is invariant under the global U($N$) 
transformation $\Phi \rightarrow U\Phi U^{\dagger}$. The
observables we shall be interested in are the singlet
combinations of closed strings of $n$ partons 
$\mbox{Tr}[\Phi({\bf x}_1),\dots,\Phi({\bf x}_n)]$, at some
fixed time. Product combinations of these closed strings of
partons (corresponding to multi-string states) are decoupled 
from the analysis in the limit $N \rightarrow \infty$,
since the quantity $1/N$ plays the role of a string coupling
constant \cite{kleb1}. In what follows, we will 
only consider the large-$N$ limit
of this model.

In two dimensions ($D=2$), the
method of Discretized Light-Cone Quantization \cite{dlcq}
has been applied in various non-perturbative 
investigations of this model, especially in connection with
a proposed $c > 1$ non-critical  string theory \cite{kleb1,ad1}.
In the present work, we will focus on the range 
$3 \leq D \leq 6$, in which transverse dynamics now play a role. 
These additional transverse degrees of freedom 
introduce ultraviolet divergences in the 
light-cone Hamiltonian, which will have to be cancelled
by some appropriate renormalization scheme. 
Various non-perturbative 
schemes for renormalizing the light-cone Hamiltonian 
can be found in the literature \cite{wilson,wegner,hiller}, but since
we will limit ourselves to a perturbative analysis, we choose to 
adopt an alternative (and possibly simpler)
strategy that yields results
in agreement with the usual (perturbative) method of covariant
renormalization. 

Working in the light-cone coordinate frame 
\begin{eqnarray}
 x^+ & = & \frac{1}{\sqrt{2}}(x^0 + x^{D-1}), \hspace{10mm}
\mbox{``time coordinate''} \\
 x^- & = & \frac{1}{\sqrt{2}}(x^0 - x^{D-1}), \hspace{10mm}
\mbox{``longitudinal space coordinate''}  \\ 
 {\bf x}^{\perp} & = & (x^1,\dots,x^{D-2}), 
\hspace{10mm} \mbox{``transverse coordinates''} 
\end{eqnarray}
one may derive from the light-cone energy-momentum tensor 
expressions for the light-cone Hamiltonian $P^-$ and
conserved total momenta $(P^+, {\bf P}^{\perp})$: 
\begin{eqnarray}
P^+ & = & \int dx^- d{\bf x}^{\perp} \hspace{1mm}
 \mbox{Tr}\left[ (\partial_- \Phi)^2 \right],
\hspace{50mm} \mbox{``longitudinal momentum''} 
\label{Pminus} \\
{\bf P}^{\perp} & = & \int dx^- d{\bf x}^{\perp} \hspace{1mm}
 \mbox{Tr}\left[ \partial_- \Phi \cdot \mbox{ \boldmath $\partial$}^{\perp}
 \Phi \right], \hspace{45mm} \mbox{``transverse momentum''}
\label{Ptrans} \\
P^- & = & \int dx^- d{\bf x}^{\perp} \hspace{1mm}
 \mbox{Tr}\left[ \frac{1}{2} m^2 \Phi^2 - \frac{1}{2}
 \mbox{ \boldmath $\partial$}_{\perp} 
 \Phi \cdot \mbox{ \boldmath $\partial$}^{\perp}
 \Phi - \frac{\lambda}{3 \sqrt{N}} \Phi^3 \right],
\hspace{4mm} \mbox{``light-cone Hamiltonian''} 
 \label{Pplus} 
\end{eqnarray}
where we adopt the notation $\mbox{ \boldmath $\partial$}_{\perp}
\Phi  \equiv (\partial_1 \Phi,\dots , \partial_{D-2} \Phi)$. 
Classically, the connection
between the light-cone field quantities $P^{\pm},{\bf P}^{\perp}$
and the Lorentz-invariant mass $M$ of the system is described by
the mass-shell relation $2P^+ P^- - |{\bf P}^{\perp}|^2 = M^2$.
In the associated quantized field theory, $P^{\pm}$ and ${\bf P}^{\perp}$
are operators, and solving the quantum field theory will mean
finding eigenstates $|\Psi>$ of the invariant mass operator
${\hat M}^2 \equiv 2P^+ P^- - |{\bf P}^{\perp}|^2$.   
The transverse momentum ${\bf P}^{\perp}$ is kinematical, and so
we may choose a reference frame in which it vanishes. The
longitudinal momentum $P^+$ is also kinematical, 
and commutes with $P^-$, so
we may (naively) formulate the bound state problem 
in terms of the light-cone Schr\"{o}dinger equation 
\begin{equation}
       2P^+ P^- |\Psi> = M^2 |\Psi>. \label{mass-shell}
\end{equation}
Of course, the above eigen-equation is meaningful only if 
the light-cone Hamiltonian $P^-$ can be shown to be well-defined
and finite. In two dimensions, the matrix elements $<\Psi_1|P^-|\Psi_2>$
are always finite, and so equation (\ref{mass-shell}) is indeed a 
{\em non-perturbative} formulation of the ($D=2$) $\Phi^3$  matrix 
theory\footnote{There is still a possibility of negative and/or
infinite ground state
energies when solving the associated infinitely-coupled set of integral 
equations. In two dimensions, this occurs for sufficiently
strong coupling \cite{kleb1,ad1}.}. 
In higher dimensions,
one expects ultraviolet divergences in 
the definition of $P^-$, 
arising from momentum integrations in the transverse sector of the theory,
so an appropriate renormalization of the light-cone Hamiltonian will
be required. If this renormalization
procedure is performed perturbatively, then we may restore 
the meaningfulness of equation (\ref{mass-shell}) only as a 
{\em perturbative} formulation of the bound state problem.   

\section{Light-Cone Quantization of $\Phi^3$ Matrix Theory} \label{lcq}
Light-cone quantization of the $\Phi^3$ matrix model is performed in 
the usual way -- namely, we impose commutation relations 
at some fixed light-cone time ($x^+ = 0$, say):
\begin{equation}
[\Phi_{ij}(x^-,{\bf x}^{\perp}), \partial_- \Phi_{lk}(y^-,{\bf y}^{\perp})
 ] = \frac{{\rm i}}{2} \delta_{i k} \delta_{j l} \delta (x^- - y^-)
 \delta ( {\bf x}^{\perp} -  {\bf y}^{\perp} ).
\label{commphi3}
\end{equation} 
The light-cone Hamiltonian $P^-$ propagates a given field configuration
in light-cone time $x^+$ while preserving this quantization
condition. At fixed  $x^+ = 0$, the
Fourier representation\footnote{
Here, the quantum conjugation ${}^{\dagger}$ does not act on indices.}  
 \begin{eqnarray}
\lefteqn{\Phi_{ij}(x^-,{\bf x}^{\perp})  =  
\frac{1}{(\sqrt{2\pi})^{D-1}} \int_0^{\infty} \frac{dk^+}{\sqrt{2 k^+}}
\int d{\bf k}^{\perp}  \times } \nonumber \\
& & \left[ 
 a_{ij}(k^+,{\bf k}^{\perp})e^{-{\rm i}(k^+ x^- - {\bf k}^{\perp} \cdot
 {\bf x}^{\perp})} +
 a_{ji}^{\dagger}(k^+,{\bf k}^{\perp})e^{+{\rm i}
(k^+ x^- - {\bf k}^{\perp} \cdot
 {\bf x}^{\perp})} \right], \label{phi}
\end{eqnarray}
together with the quantization condition (\ref{commphi3}),
imply the relation 
\begin{equation}
[ a_{i j}(k^+,{\bf k}^{\perp}), 
a_{k l}^{\dagger}({\tilde k}^+,{\tilde {\bf k}}^{\perp})] = 
\delta_{i k} \delta_{j l} \delta (k^+ - {\tilde k}^+)
 \delta ( {\bf k}^{\perp} -  {\tilde {\bf k}}^{\perp} ).
\end{equation}
It is now a matter of substituting the Fourier representation
(\ref{phi}) for the quantized matrix field $\Phi$ into definitions
(\ref{Pminus}),(\ref{Ptrans}), and (\ref{Pplus}), to obtain
the following quantized expressions for the light-cone
Hamiltonian and conserved total momenta:
\begin{eqnarray}
\lefteqn{ : P^+ : \hspace{3mm}  =  \int_0^{\infty} dk^+ 
    \int d{\bf k}^{\perp} 
           \hspace{1mm} k^+ \cdot a_{i j}^{\dagger}(k^+,{\bf k}^{\perp})
           a_{i j}(k^+,{\bf k}^{\perp}), } \\
\lefteqn{ : {\bf P}^{\perp} : 
\hspace{3mm}  =  \int_0^{\infty} dk^+ 
     \int d{\bf k}^{\perp} \hspace{1mm} {\bf k}^{\perp}
\cdot
           a_{i j}^{\dagger}(k^+,{\bf k}^{\perp})
           a_{i j}(k^+,{\bf k}^{\perp}), } \\
\lefteqn{ :P^-: \hspace{3mm}  =  \int_0^{\infty} dk^+ \int d{\bf k}^{\perp} 
            \left( \frac{m^2 + |{\bf k}^{\perp}|^2}{2 k^+} \right)
           a_{i j}^{\dagger}(k^+,{\bf k}^{\perp})
           a_{i j}(k^+,{\bf k}^{\perp}) } \nonumber \\
        & - & \lambda \frac{2^{-3/2} N^{-1/2}}
{ (\sqrt{2 \pi})^{D-1}
                } \int_0^{\infty} \frac{
 dk_1^+ dk_2^+ dk_3^+ }{\sqrt{k_1^+ k_2^+ k_3^+}} \delta (
 k_1^+ + k_2^+ - k_3^+) \int d{\bf k}_1^{\perp} 
 d{\bf k}_2^{\perp}  d{\bf k}_3^{\perp} 
 \hspace{1mm} \delta ({\bf k}_1^{\perp} +
 {\bf k}_2^{\perp}  - {\bf k}_3^{\perp}) \times \nonumber \\
& & \left[ a_{i j}^{\dagger}(k_3^+,{\bf k}_{3}^{\perp})
        a_{i k}(k_1^+,{\bf k}_{1}^{\perp})
      a_{k j}(k_2^+,{\bf k}_{2}^{\perp}) + 
        a_{i k}^{\dagger}(k_1^+,{\bf k}_{1}^{\perp})
      a_{k j}^{\dagger}(k_2^+,{\bf k}_{2}^{\perp}) 
    a_{i j}(k_3^+,{\bf k}_{3}^{\perp}) \right]
\end{eqnarray} 
(repeated indices are summed over $1$ to $N$). The light-cone vacuum
$|0>$  will be 
defined by stipulating  
\begin{equation}
          a_{i j}(k^+, {\bf k}^{\perp})\cdot |0> = 0,    \hspace{8mm}
k^+ > 0,
\end{equation}
for all $i,j=1,\dots ,N$, all transverse momenta ${\bf k}^{\perp}$,
and all positive longitudinal
momenta $k^+$. It should be stressed that we will always work
in the infinite volume {\em continuum}, and so the limit $k^+ \rightarrow 0^+$
may be taken in any expression. Moreover, we may leave all field quantities
undefined at $k^+ = 0$\footnote{So-called {\em zero modes}
are  important in various non-continuum formulations
of field theories. See \cite{dave,sho}, and references therein.}. 

In this case, the vacuum is (trivially) a zero eigenstate of the 
light-cone Hamiltonian,
which is a unique feature of light-cone quantization.
The Fock space of particle states on which the light-cone Hamiltonian
acts non-trivially is generated by states of the form
$a_{i_1 j_1}^{\dagger}(k_1^+,{\bf k}_1^{\perp}) \dots 
a_{i_n j_n}^{\dagger}(k_n^+,{\bf k}_n^{\perp})|0>$, $n \geq 1$.
A subspace may be formed by considering the space generated by
the following (orthogonal) states with fixed total 
momenta $(P^+,{\bf P}^{\perp})$,
\begin{equation}
\frac{1}{\sqrt{N^n}} \mbox{Tr}[
a^{\dagger}(k_1^+,{\bf k}_1^{\perp})\dots
a^{\dagger}(k_n^+,{\bf k}_n^{\perp})|0>, \hspace{7mm} 
 \sum_{i=1}^{n}k^+_i = P^+ \hspace{3mm} 
\sum_{i=1}^{n} {\bf k}^{\perp}_i = {\bf P}^{\perp}.
\end{equation}
These states transform as singlets under the 
$\Phi \rightarrow U \Phi U^{\dagger}$
global U($N$) symmetry of the action, and
have a natural `closed string' interpretation; they
are (trivially) eigenstates of the 
momentum operators $P^+$ and ${\bf P}^{\perp}$ respectively.
Choosing a reference frame for which ${\bf P}^{\perp}$ vanishes,
the bound state formulation of the theory is given by the
light-cone Schr\"odinger equation (\ref{mass-shell}).
In the next section, we show that this is equivalent
to an infinite set of coupled integral equations.

\section{Bound State Integral Equations for $\Phi^3$ Matrix Theory} 
\label{boundstate}  
The expression for the superposition of all singlet states
with conserved total momenta $P^+$, ${\bf P}^{\perp}= {\bf 0}^{\perp}$
has the explicit form 
\begin{eqnarray}
|\Psi> & = &\sum_{n=1}^{\infty}
 \int_0^{1} dx_1^+ \dots dx_n^+ \hspace{1mm} 
 \delta (x_1^+ + \cdots + x_n^+ - 1) \int
 d{\bf k}_1^{\perp} \dots d{\bf k}_n^{\perp} \hspace{1mm}
   \delta ( {\bf k}_1^{\perp} + \cdots + {\bf k}_n^{\perp})
 \times  \nonumber \\
& & f_n({\bf k}_1,\dots,{\bf k}_n) \frac{1}{\sqrt{N^n}}
\mbox{Tr} [ a^{\dagger}({\bf k}_1) \dots  a^{\dagger}({\bf k}_n)]|0>,
\label{goof}
\end{eqnarray}
where we have introduced the longitudinal momentum fractions
$x_i = k^+_i/P^+$, and notation ${\bf k}_i \equiv (x_i, {\bf k}^{\perp}_i)$.
Note that the wavefunctions $f_n$ are defined for all 
transverse momenta ${\bf k}_i^{\perp}$,
and all momentum fractions $x_i > 0$ such that $\sum_{i=1}^n x_i = 1$.
Substituting  expression (\ref{goof}) into the light-cone Schr\"odinger
equation (\ref{mass-shell}) yields  infinitely many coupled
integral equations for the wavefunctions $f_n$, $n \geq 1$. 
For convenience, 
we choose to write these integral equations in terms of the 
modified functions $g_n({\bf k}_1,\dots,{\bf k}_n) =
f_n({\bf k}_1,\dots,{\bf k}_n)/
\sqrt{x_1 x_2 \dots x_n}$, 
$n \geq 1$: 
\begin{eqnarray}
M^2 g_n({\bf k}_1,\dots,{\bf k}_n) & = & \left( \frac{m^2 + 
|{\bf k}_1^{\perp}|^2 }{x_1} \right) g_n({\bf k}_1,\dots,{\bf k}_n) 
\nonumber \\
& - & y_D \hspace{1mm}\frac{ g_{n-1}({\bf k}_1 + {\bf k}_2, {\bf k}_3,\dots,
     {\bf k}_n)}{x_1 x_2}    \nonumber \\
& - & y_D \int_0^1 d\alpha d\beta \hspace{1mm} \delta (\alpha + \beta -1)
\int d{\bf p}^{\perp} d{\bf q}^{\perp} \hspace{1mm}
 \delta ( {\bf p}^{\perp} +  {\bf q}^{\perp} - 
 {\bf k}_1^{\perp} ) \times \nonumber \\
& & \hspace{10mm}
g_{n+1}((\alpha x_1,{\bf p}^{\perp}),
             (\beta x_1,{\bf q}^{\perp}),
              {\bf k}_2,\dots,{\bf k}_n)  \nonumber \\
& + & \mbox{  cyclic permutations of $({\bf k}_1,\dots ,{\bf k}_n)$,}
\label{inteqns}
\end{eqnarray}
where  $y_D \equiv \lambda / \sqrt{2}(\sqrt{2 \pi})^{D-1}$ is
the coupling constant in the theory. In terms of some 
mass scale $\mu$, the coupling constant has dimensions
$[y_D] = \mu^{(6-D)/2}$, and so it becomes dimensionless
in six dimensions.
 
For sufficiently high space-time dimensions, the integral equations
(\ref{inteqns}) are expected to possess ultraviolet divergences
arising from  integrations over transverse momentum space. 
Consequently, the equations are 
only mathematically meaningful
if we introduce some ultraviolet cut-off, and physically 
acceptable if the bound state masses $M^2$ are finite and 
convergent as this cut-off is sent to infinity.   

One way to study the ultraviolet properties of the theory
is to consider the effect of
taking the limit $x_1 \rightarrow 0^+$. 
Notice that the kinetic mass term on the right-hand side of 
(\ref{inteqns}) possesses a  
$\frac{1}{x_1}$ pole, and so it diverges when 
$x_1 \rightarrow 0^+$.
If $M^2$ is finite, 
the term on the left-hand side cannot compensate for this 
singularity, so we must have
a special cancelation condition between terms on the right-hand
side as $x_1$ is made to vanish. To deduce the precise form of
this relation, simply multiply both sides of (\ref{inteqns})
by $x_1$, and then take the limit $x_1 \rightarrow 0^+$. 
The result is $(n \geq 2)$:
\begin{eqnarray}
 \lefteqn{ (m^2 + |{\bf k}_1^{\perp}|^2) \cdot
 g_n((0^+,{\bf k}_1^{\perp}), {\bf k}_2,\dots,{\bf k}_n)  = }\nonumber \\
& &  y_D \left[ \frac{g_{n-1}((0^+,{\bf k}_1^{\perp}) + {\bf k}_2,
     {\bf k}_3,\dots, {\bf k}_n)}{x_2} +
         \frac{g_{n-1}({\bf k}_2,\dots,
     {\bf k}_{n-1},{\bf k}_n + ( 0^+,{\bf k}_1^{\perp}))}{x_n} \right] 
\nonumber \\ 
& + & y_D \int_0^{\infty} d\alpha d\beta \hspace{1mm} 
 \delta ( \alpha + \beta - 1 ) \int
 d{\bf p}^{\perp} d{\bf q}^{\perp} \hspace{1mm} 
 \delta ({\bf p}^{\perp}  + {\bf q}^{\perp} - {\bf k}_1^{\perp} )   
\times \nonumber \\
& & \hspace{25mm} \lim_{x_1 \rightarrow 0^+}
     x_1 \cdot g_{n+1}((\alpha x_1, {\bf p}^{\perp}),
                       (\beta x_1, {\bf q}^{\perp}),
                       {\bf k}_2,\dots,{\bf k}_n),
\label{firstladder}
\end{eqnarray}
where ${\bf k}_i \equiv (x_i,{\bf k}_i^{\perp})$ as usual, 
and $(0^+,{\bf k}_i^{\perp}) \equiv
\lim_{x_i \rightarrow 0^+} {\bf k}_i$. 
Of course, we obtain entirely analogous results 
for  each limit $x_i \rightarrow 0^+$.
The importance of such `ladder relations' was 
first recognized in a light-cone study of dimensionally
reduced QCD \cite{coll}, and in subsequent work
on meson wavefunctions from large $N_c$ light-front ${\mbox{QCD}}_{3+1}$
 \cite{abd}. 
Recent detailed studies  have shown that it is
possible to systematically eliminate the integral
terms appearing in such relations in favour of
renormalizing  existing terms, and introducing
other integrals involving wavefunctions from
higher Fock sectors \cite{dall2,houch}. We shall be exploiting these
ideas in what follows.      
  
A nice property of relation (\ref{firstladder}) is the absence
of the mass 
eigenvalue $M^2$, since it decouples in the limit $x_1 \rightarrow 0^+$
in equations (\ref{inteqns}). Note also that
the wavefunction in the integral involves a limit where 
{\em two} longitudinal momenta are simultaneously sent to zero.
This is the basis of a recursive scheme, since we may 
always use equations (\ref{inteqns}) to express a wavefunction
with $m$ simultaneously vanishing longitudinal momenta in terms of
one with $m+1$ and one with $m-1$ vanishing  
momenta (the mass eigenvalue $M^2$ always decouples in such
relations). A recursive scheme then enables one to eliminate 
wavefunctions from a given Fock sector in favour of
renormalizing lower
Fock sector components, and introducing higher
Fock sector dependent integral expressions. 
If one imposes a Tamm-Dancoff truncation \cite{tammy} (i.e. if
one assumes $g_n \equiv 0$ for $n$ greater than some fixed
integer) then this recursive procedure will terminate, and
all existing terms in the ladder relations will receive no further
corrections. 
In this sense, Tamm-Dancoff truncation is simply perturbation
theory. More will be said on this issue in Section \ref{tamm}.   

To elucidate on the above-mentioned recursion procedure,
we begin by analyzing in more detail the integral 
appearing in relation (\ref{firstladder}). 
Firstly, in equation (\ref{inteqns}),
make the substitutions $n \rightarrow n+1$, ${\bf k}_1 
\rightarrow (\alpha x_1, {\bf p}^{\perp})$,
 ${\bf k}_2 
\rightarrow (\beta x_1, {\bf q}^{\perp})$, 
 multiply both
sides by $x_1^2$, and
then let $x_1 \rightarrow 0^+$. We end up with 
\begin{eqnarray}
\lefteqn{ \left( \frac{m^2 + |{\bf p}^{\perp}|^2}{\alpha}
           + \frac{m^2 + |{\bf q}^{\perp}|^2}{\beta} \right)
    \lim_{x_1 \rightarrow 0^+} 
        x_1 \cdot g_{n+1}((\alpha x_1, {\bf p}^{\perp}),
                       (\beta x_1, {\bf q}^{\perp}),
                       {\bf k}_2,\dots,{\bf k}_n) = } \nonumber \\
& & y_D\cdot \frac{1}{\alpha \beta} \cdot \lim_{x_1 \rightarrow 0^+}
 g_n(((\alpha + \beta)x_1,{\bf p}^{\perp} + {\bf q}^{\perp}),
 {\bf k}_2,\dots {\bf k}_n) \nonumber \\
& + & y_D \int_0^1 d{\tilde \alpha} d{\tilde \beta} \hspace{1mm}
 \delta ({\tilde \alpha} +  {\tilde \beta} - 1) \int
 d{\tilde {\bf p}}^{\perp}  d{\tilde {\bf q}}^{\perp}
\hspace{1mm} \delta ({\tilde {\bf p}}^{\perp} + {\tilde {\bf q}}^{\perp}
 - {\bf p}^{\perp}) \times \nonumber \\
& & \hspace{25mm}
\lim_{x_1 \rightarrow 0^+} x_1^2 \cdot 
g_{n+2}(({\tilde \alpha} \alpha x_1, {\tilde {\bf p}}^{\perp}),
        ({\tilde \beta}  \alpha x_1, {\tilde {\bf q}}^{\perp}),
        (\beta x_1, {\bf q}^{\perp}),{\bf k}_2,\dots {\bf k}_n) \nonumber \\
& + &  y_D \int_0^1 d{\tilde \alpha} d{\tilde \beta} \hspace{1mm}
 \delta ({\tilde \alpha} +  {\tilde \beta} - 1) \int
 d{\tilde {\bf p}}^{\perp}  d{\tilde {\bf q}}^{\perp}
\hspace{1mm} \delta ({\tilde {\bf p}}^{\perp} + {\tilde {\bf q}}^{\perp}
 - {\bf q}^{\perp}) \times \nonumber \\
& & \hspace{25mm}
\lim_{x_1 \rightarrow 0^+} x_1^2 \cdot 
g_{n+2}((\alpha x_1, {\bf p}^{\perp}),
        ({\tilde \alpha} \beta x_1, {\tilde {\bf p}}^{\perp} ),
        ({\tilde \beta} \beta x_1, {\tilde {\bf q}}^{\perp} ),  
{\bf k}_2,\dots {\bf k}_n). 
\label{twozeros}
\end{eqnarray}  
We may now use this last relation 
to re-express the integral appearing in (\ref{firstladder}).
The new (yet equivalent) ladder relation we obtain after 
making this substitution is  
\begin{eqnarray}
 \lefteqn{ \left[ 
m^2 + |{\bf k}_1^{\perp}|^2 - y_D^2 \Gamma_D({\bf k}_1^{\perp}) \right] 
\cdot
 g_n((0^+,{\bf k}_1^{\perp}), {\bf k}_2,\dots,{\bf k}_n)  = }\nonumber \\
& &  y_D \left[ \frac{g_{n-1}((0^+,{\bf k}_1^{\perp}) + {\bf k}_2,
     {\bf k}_3,\dots, {\bf k}_n)}{x_2} +
         \frac{g_{n-1}({\bf k}_2,\dots,
     {\bf k}_{n-1},{\bf k}_n + ( 0^+,{\bf k}_1^{\perp}))}{x_n} \right] 
\nonumber \\
& + &  y_D \cdot {\cal I}_{n+2}({\bf k}_1^{\perp}), 
\label{laddersecond}
\end{eqnarray}
where 
\begin{equation}
\Gamma_D({\bf k}_1^{\perp}) = 
\int_0^1 d\alpha d\beta \hspace{1mm}
\delta (\alpha + \beta - 1)
\int d{\bf p}^{\perp} d{\bf q}^{\perp} \hspace{1mm}  
 \delta ({\bf p}^{\perp} + {\bf q}^{\perp} - {\bf k}_1^{\perp} )
 \frac{1}{
\beta (m^2 + |{\bf p}^{\perp}|^2) +
\alpha (m^2 + |{\bf q}^{\perp}|^2 )},
\label{gammak1}
\end{equation}
and
\begin{eqnarray}
\lefteqn{ {\cal I}_{n+2}({\bf k}_1^{\perp}) = } \nonumber \\
& &  
\int_0^1 d\alpha d\beta \hspace{1mm}
\delta (\alpha + \beta - 1) \cdot \alpha \beta \cdot
\int_0^1 d{\tilde \alpha} d{\tilde \beta} \hspace{1mm}
\delta ({\tilde \alpha} + {\tilde \beta} - 1) \times \nonumber \\
& & 
\int d{\bf p}^{\perp} d{\bf q}^{\perp} \hspace{1mm}  
 \delta ({\bf p}^{\perp} + {\bf q}^{\perp} - {\bf k}_1^{\perp} )
 \frac{1}{
\beta (m^2 + |{\bf p}^{\perp}|^2) +
\alpha (m^2 + |{\bf q}^{\perp}|^2 )} 
\int
 d{\tilde {\bf p}}^{\perp}  d{\tilde {\bf q}}^{\perp}
\hspace{1mm} \delta ({\tilde {\bf p}}^{\perp} + {\tilde {\bf q}}^{\perp}
 - {\bf p}^{\perp}) \times \nonumber \\
& & \hspace{25mm}
\lim_{x_1 \rightarrow 0^+} x_1^2 \cdot 
g_{n+2}(({\tilde \alpha} \alpha x_1, {\tilde {\bf p}}^{\perp}),
        ({\tilde \beta}  \alpha x_1, {\tilde {\bf q}}^{\perp}),
        (\beta x_1, {\bf q}^{\perp}),{\bf k}_2,\dots {\bf k}_n) \nonumber \\
&+& \int_0^1 d\alpha d\beta \hspace{1mm}
\delta (\alpha + \beta - 1) \cdot \alpha \beta \cdot
\int_0^1 d{\tilde \alpha} d{\tilde \beta} \hspace{1mm}
\delta ({\tilde \alpha} + {\tilde \beta} - 1) \times \nonumber \\
& & 
\int d{\bf p}^{\perp} d{\bf q}^{\perp} \hspace{1mm}  
 \delta ({\bf p}^{\perp} + {\bf q}^{\perp} - {\bf k}_1^{\perp} )
 \frac{1}{
\beta (m^2 + |{\bf p}^{\perp}|^2) +
\alpha (m^2 + |{\bf q}^{\perp}|^2 )} 
\int
 d{\tilde {\bf p}}^{\perp}  d{\tilde {\bf q}}^{\perp}
\hspace{1mm} \delta ({\tilde {\bf p}}^{\perp} + {\tilde {\bf q}}^{\perp}
 - {\bf q}^{\perp}) \times \nonumber \\
& & \hspace{25mm}
\lim_{x_1 \rightarrow 0^+} x_1^2 \cdot 
g_{n+2}((\alpha x_1, {\bf p}^{\perp}),
        ({\tilde \alpha} \beta x_1, {\tilde {\bf p}}^{\perp} ),
        ({\tilde \beta} \beta x_1, {\tilde {\bf q}}^{\perp} ),  
{\bf k}_2,\dots {\bf k}_n).
\end{eqnarray}
In words, we eliminate the dependence 
on the $n+1$ parton wavefunction $g_{n+1}$ appearing
in the integral of relation (\ref{firstladder})
in favour of 
renormalizing 
the kinetic mass term 
$(m^2 + |{\bf k}_1^{\perp}|^2) \rightarrow
 [ m^2 + |{\bf k}_1^{\perp}|^2 -y_D^2 \Gamma_D({\bf k}_1^{\perp})]$,
and introducing
further (more complicated) integral expressions (summarily denoted by
${\cal I}_{n+2}$) 
involving $n+2$ parton wavefunctions
$g_{n+2}$. The wavefunctions appearing in these integrals now
have {\em three} simultaneously vanishing longitudinal momenta.
Of course, a small-$x$ analysis of (\ref{inteqns}) allows the
wavefunction $g_{n+2}$ with three  vanishing
momenta to be re-expressed in terms of 
$n+1$ parton  wavefunctions $g_{n+1}$ (with two vanishing momenta)
and $n+3$ parton wavefunctions $g_{n+3}$ (four vanishing momenta). 
Repeated application of
this procedure will give further (higher order) corrections
to the kinetic mass term, in even powers of the coupling $y_D$.
If we impose a Tamm-Dancoff truncation on the Fock space of states,
then eventually this recursive procedure will end, and the 
kinetic mass term will receive no further corrections. 

At this point, no comment has been made about the 
ultraviolet properties of the kinetic mass renormalization
$\Gamma_D({\bf k}_1^{\perp})$ defined by equation
$(\ref{gammak1})$. We address this topic next,
which will point the way towards a
practical scheme for renormalizing
the light-cone Hamiltonian required in numerical approaches.

\section{Renormalization of $\Phi^3$ Matrix Theory}
\label{renorm}
Let us now consider evaluating the integral appearing
in the definition of $\Gamma_D({\bf k}_1^{\perp})$ for
the kinetic mass renormalization. 
Our analysis begins in $2+1$ space-time dimensions.

{\em Case 1: D=3}

For the simplest case $D=3$, there is only one transverse coordinate,
and so the integral in (\ref{gammak1}) is one dimensional. In fact,
it may be integrated exactly to give the formula
\begin{eqnarray}
\Gamma_{D=3} ({\bf k}_1^{\perp}) & = & \frac{2\pi}{|{\bf k}_1^{\perp}|}
 \tan^{-1} \frac{ |{\bf k}_1^{\perp}|}{2 m} \nonumber \\
& = & \frac{\pi}{m} \left[ 1 - \frac{ |{\bf k}_1^{\perp}|^2}{12 m^2}
 +    \frac{ |{\bf k}_1^{\perp}|^4}{80 m^4} +\hspace{3mm} \cdots \right].
\label{gamma3d}
\end{eqnarray}
Note that $\Gamma_{D=3}({\bf k}_1^{\perp})$ is finite,
so we are not forced to renormalize the light-cone
Hamiltonian in $2+1$ dimensions at this leading order.
We also see from (\ref{gamma3d}) above that 
$\Gamma_{D=3}({\bf k}_1^{\perp})$ contains
transverse interactions beyond 
$|{\bf k}_1^{\perp}|^2$. In this sense,
we have obtained information about the {\em effective} light-cone
Hamiltonian, since, at this leading order, such
additional  interactions
are meant to compensate for the absence of 
explicit higher Fock sector
dependence.

Note finally that if one were able to show that 
all higher order corrections involve only ultraviolet finite integrals,
then this would uphold the claim that 
the bound state
integral equations (\ref{inteqns}) represent a  
non-perturbative
formulation of $\Phi^3$ matrix theory in $D=3$ space-time
dimensions.  

{\em Case 2: D=4,5 and 6}

In space-time dimensions $D \geq 4$, the 
evaluation of the integral $(\ref{gammak1})$
does not appear to admit any simple closed analytical
expression. However,
we do have a series expansion in powers of 
$|{\bf k}_1^{\perp}|$:
\begin{eqnarray}
\Gamma_D({\bf k}_1^{\perp}) & = & \int d{\bf p}^{\perp} \left\{
 \frac{1}{m^2 + |{\bf p}^{\perp}|^2} 
+ \frac{|{\bf p}^{\perp}|\cos \theta}
       {(m^2 + |{\bf p}^{\perp}|^2)^2}
|{\bf k}_1^{\perp}| +
\frac{-3m^2 +  |{\bf p}^{\perp}|^2 (1+4 \cos 2\theta)}
{ 6(m^2 + |{\bf p}^{\perp}|^2)^3}|{\bf k}_1^{\perp}|^2 
\right. \nonumber \\
& + &  \frac{|{\bf p}^{\perp}|\cos \theta 
( -4m^2 - |{\bf p}^{\perp}|^2(1 - 3 \cos 2\theta ))}
{3 (m^2 + |{\bf p}^{\perp}|^2)^4} |{\bf k}_1^{\perp}|^3 \nonumber \\
& + & \left. \frac{10 m^4 -5 m^2 |{\bf p}^{\perp}|^2 
(5+9 \cos 2\theta ) +   |{\bf p}^{\perp}|^4
(1 + 3 \cos 2\theta + 12 \cos 4\theta)}{30 (m^2 + |{\bf p}^{\perp}|^2)^5}
 |{\bf k}_1^{\perp}|^4 \hspace{3mm} + \hspace{3mm} \cdots \frac{}{} \right\},
\end{eqnarray}
where $\theta$ is the angle between ${\bf k}_1^{\perp}$ and
${\bf p}^{\perp}$. Of course, we may always rotate the space of
integration so that $\theta$ is one of the polar angles
in a polar co-ordinate representation of the underlying $D-2$ dimensional
space of integration. In such a representation,
we have $d{\bf p}^{\perp}  = \rho^{D-3} d\rho d\Omega$, where
$\rho = |{\bf p}^{\perp}|$, and $d \Omega$ denotes integration over
the polar angles (there are $D-3$ of them).

Applying these ideas in $D=4,5$ and $6$ space-time dimensions, we    
deduce the following expansions:    
\begin{eqnarray}
\Gamma_{D=4}({\bf k}_1^{\perp}) & = &
 \left( \pi \log\frac{m^2 + \Lambda_{\perp}^2}{m^2} \right)
  + \left( -\frac{\pi}{6 m^2}  + \frac{\pi m^2}{
 3(m^2 + \Lambda_{\perp}^2)^2} - \frac{\pi}{6(m^2 + \Lambda_{\perp}^2)}
 \right)
 |{\bf k}_1^{\perp}|^2 \hspace{3mm} + \hspace{3mm} \cdots \\
 \Gamma_{D=5}({\bf k}_1^{\perp}) & = &
\left( 4 \pi  \Lambda_{\perp} - 4 \pi m \tan^{-1} \frac{\Lambda_{\perp}}{m} 
 \right) +
 \left( \frac{ 2 \Lambda_{\perp} m^2 \pi}{3(m^2+\Lambda_{\perp}^2)^2}
 - \frac{2 \Lambda_{\perp} \pi}{3 (m^2+\Lambda_{\perp}^2)}
\right) |{\bf k}_1^{\perp}|^2 \hspace{3mm} + \hspace{3mm} \cdots \\
 \Gamma_{D=6}({\bf k}_1^{\perp}) & = &
\left( \pi^2 \Lambda_{\perp}^2 - \pi^2 m^2 
\log\frac{m^2 + \Lambda_{\perp}^2}{m^2} \right) \nonumber \\
& & 
+ \left( \frac{\pi^2}{6}\log\frac{m^2 + \Lambda_{\perp}^2}{m^2}
 - \frac{\pi^2}{2} + \frac{5 \pi^2 m^2}{6(m^2+\Lambda_{\perp}^2)}
 - \frac{\pi^2 m^4}{3(m^2+\Lambda_{\perp}^2)^2} \right)
|{\bf k}_1^{\perp}|^2 \hspace{3mm} + \hspace{3mm} \cdots \label{gamma6} 
\end{eqnarray}         
where we have adopted the cut-off scheme 
$|{\bf p}^{\perp}| \leq \Lambda_{\perp}$ in the evaluation of the integrals.
Note that only even powers of
$|{\bf k}_1^{\perp}|$ appear in the
expansion. Moreover, it can easily be shown (counting arguments)
that all terms involving quartic
or higher powers of $|{\bf k}_1^{\perp}|$ are finite as the cut-off 
$\Lambda_{\perp}$ is sent to infinity. 

It is clear that for $D \geq 4$, constituent masses receive
diverging corrections from the leading term in the 
$\Gamma_D({\bf k}_1^{\perp})$ expansion in the 
ultraviolet limit $\Lambda_{\perp}  \rightarrow \infty$.
Recall that our objective is to ensure that the light-cone Hamiltonian
$P^-$ is well defined and finite in this ultraviolet limit.
Consequently, subtracting  infinities 
which appear in relations such as 
(\ref{laddersecond}), which are derived from the Hamiltonian
in a special limit,
yield only {\em necessary}
`finiteness conditions' for the full light-cone Hamiltonian. 
Even after implementing these conditions (say, at all
orders), there is still the possibility of a negative and/or
divergent ground state eigenvalue for $M^2$ after 
solving the infinitely coupled set of (renormalized) 
equations appearing
in (\ref{inteqns}).  

With this caveat in mind, we may now briefly
discuss the 
formal procedure of perturbative renormalization
which is applicable to the $\Phi^3$ matrix model.
Recall that $\Gamma_D({\bf k}_1^{\perp})$ appears in the
modified relation (\ref{laddersecond}), and renormalizes
the kinetic mass term  $(m^2+|{\bf k}_1^{\perp}|^2)
\rightarrow [m^2+|{\bf k}_1^{\perp}|^2 -y_D^2 \Gamma_D({\bf k}_1^{\perp})]$. 
Finiteness of this relation
requires that any ultraviolet divergence appearing in 
$\Gamma_D({\bf k}_1^{\perp})$ must be subtracted by an appropriate
re-definition of coupling constants. 
For $D=4$ and $D=5$, only the constant term in the expansion
for $\Gamma_D({\bf k}_1^{\perp})$ diverges as $\Lambda_{\perp}
\rightarrow \infty $, and so  
eliminating these infinities is easily accomplished by 
introducing a cut-off dependent mass parameter $m=m(\Lambda_{\perp})$.
We omit details.
The expansion for $\Gamma_D({\bf k}_1^{\perp})$ also
introduces higher powers of $|{\bf k}_1^{\perp}|$, which
are finite in the ultraviolet limit. Consequently, in 
four and five space-time dimensions, 
effective transverse interactions are present in a
leading order analysis. 

Turning our attention to six space-time dimensions, $D=6$,
it is evident from (\ref{gamma6}) that the coefficient 
of $|{\bf k}_1^{\perp}|^2$ diverges logarithmically,
so an appropriate renormalization will be required. In
this case, there is no bare coupling in the original Lagrangian
that is capable of absorbing this divergence, but we do
have the freedom to renormalize the wavefunctions $g_n$
by some  multiplicative factor. In the present
case, we divide both sides of equation (\ref{laddersecond})
by the 
(divergent) coefficient of  $|{\bf k}_1^{\perp}|^2$
in the renormalized kinetic term
$[ m^2 +|{\bf k}_1^{\perp}|^2 - y_6^2 \Gamma_{D=6}({\bf k}_1^{\perp})]$,
 which is equivalent 
to a multiplicative renormalization of the wavefunction.
This modified form of relation  (\ref{laddersecond}) is now 
evidently free of any ultraviolet divergences if we can choose
cut-off dependent coupling constants $m=m(\Lambda_{\perp})$
and $y_6 = y_6(\Lambda_{\perp})$  that satisfy a coupled
set of equations:
\begin{eqnarray}
 m_0^2 & = & \frac{
 \left[ m^2 - y_6^2\left( 
\pi^2 \Lambda_{\perp}^2 - \pi^2 m^2 
\log\frac{m^2 + \Lambda_{\perp}^2}{m^2} \right) \right]}{
1 - y_6^2 \left( 
  \frac{\pi^2}{6}\log\frac{m^2 + \Lambda_{\perp}^2}{m^2}
 - \frac{\pi^2}{2} + \frac{5 \pi^2 m^2}{6(m^2+\Lambda_{\perp}^2)}
 - \frac{\pi^2 m^4}{3(m^2+\Lambda_{\perp}^2)^2} \right) } \\
 y_0 & = & \frac{ y_6 }{ 
1 - y_6^2 \left( 
  \frac{\pi^2}{6}\log\frac{m^2 + \Lambda_{\perp}^2}{m^2}
 - \frac{\pi^2}{2} + \frac{5 \pi^2 m^2}{6(m^2+\Lambda_{\perp}^2)}
 - \frac{\pi^2 m^4}{3(m^2+\Lambda_{\perp}^2)^2} \right) }
\label{sec}
\end{eqnarray}
where $m_0$ and $y_0$ are arbitrary parameters which are 
fixed and finite. A formal perturbative renormalization
procedure simply involves finding solutions up to 
order $y_6^2$ i.e. calculating the leading perturbative
correction to the zero order solution $m=m_0$, $y_6=y_0$.
These results are  entirely consistent with
the usual covariant approach to perturbative renormalization,
applied to  $\Phi^3$ scalar theory. Interestingly, in the 
light-cone approach we have adopted here, 
a Wick rotation and/or introduction of an
${\rm i}\epsilon$ pole was not necessary in the calculations.

{\em Remark:} In the previous discussion, we assumed
$\Lambda_{\perp}/m(\Lambda_{\perp}) \rightarrow \infty$ in
the ultraviolet limit. However, consider the ansatz
$m(\Lambda_{\perp}) = c \cdot \Lambda_{\perp}$, for some non-zero
constant $c$. Then the coefficient of $|{\bf k}_1^{\perp}|^2$
in the expansion for $\Gamma_{D=6}({\bf k}_1^{\perp})$ is
finite! The (renormalized) mass term, however, has the form
\begin{equation}
     \left[ c^2 - y_6^2 \left( 
   \pi^2 - c^2 \pi^2 \log \frac{1+c^2}{c^2}
        \right) \right] \cdot \Lambda_{\perp}^2
\end{equation}
and so we avoid any ultraviolet divergences only if
we choose $c$ such that the above term vanishes\footnote{
A real solution for $c$ exists if $y_6 \neq 0$.}.
There is no wavefunction renormalization (at this order),
and so the coupling constant $y_6$ appearing above is the bare
one.
We conclude, therefore, that in the special case of
$D=6$ dimensions, only mass renormalization 
is required (at leading order)
if we allow the renormalized mass to vanish. 

\section{Renormalization in Numerical Approaches}
\label{tamm} 
The renormalization procedure outlined in the
previous section requires some important modifications
if we wish to consider diagonalizing
the light-cone Hamiltonian 
numerically via Discretized Light-Cone
Quantization methods. 

Firstly, any numerical procedure will be forced to
impose some truncation of the Fock space of states;
in particular, in the bound state equations (\ref{inteqns}),
we simply neglect wavefunctions $g_n$ for $n$ larger
than some fixed integer - $N_0$, say. This has 
an immediate effect on the  
form of the ladder relations: For
$n=N_0$, there is no integral appearing in relation
$(\ref{firstladder})$, and so any ultraviolet divergences
are absent. Therefore, at the $n=N_0$ level of
the bound state equations, there is no need for
any renormalization of the couplings. However, 
for $n=N_0 -1$, the integral term
appearing in (\ref{firstladder}) will yield ultraviolet
divergences, but only at leading order, since now
${\cal I}_{n+2}$ in relation (\ref{laddersecond})
vanishes. Similarly, for $n=N_0 -2$, equations 
(\ref{inteqns}) will require renormalizations
up to next-to-leading order only. It follows that
the kinetic mass term 
for the (lowest) wavefunction $g_1$ will require the highest
order corrections.  
The conclusion is that
the renormalization procedure is {\em Fock sector dependent},
and that Tamm-Dancoff truncation is essentially
perturbation theory (in this approach).

\section{Discussion}
\label{the-end}
The content of this work represents a 
preliminary study on the renormalizability of matrix field 
theories in $D \geq 3$ dimensions
which are quantized on the light-cone\footnote{ 
A study of finiteness and renormalizability of
light-cone Hamiltonians in $D=2$ dimensions can
be found in \cite{mat}}.
No reference
to the equal-time quantization formalism has been made;
rather, we have pursued a strategy that relies exclusively
on the special properties of field theories formulated
in the light-cone frame. In particular, we observed 
that it is possible to
study the  ultraviolet properties of the theory -- order 
by order in powers of the coupling --
by considering regions of phase space where one or more longitudinal
momenta are made to vanish. Our formalism also
provides information on effective interactions 
which are needed to compensate for the elimination of
higher Fock sector components. These results are 
a precise realization of the commonly held belief
that the dynamics at small momentum fraction-$x$ 
must be intimately linked
with high energy dynamics in the transverse sector of the theory.

One interesting feature in this formalism is that
one obtains complete agreement with the usual 
covariant approach to perturbative renormalization. However,
in the light-cone 
approach adopted here,
there is no need to introduce any Wick rotation and/or
${\rm i}\epsilon$ pole to regulate spurious divergences.
Moreover, no knowledge of Feynman diagram techniques is required.
In this respect, the concept of renormalizability of field theories
in Minkowski space has a firmer -- if
not completely unambiguous -- foundation 
and interpretation in the light-cone coordinate frame.

An immediate consequence of this is that 
one can determine practical renormalization
schemes for regulating the light-cone Hamiltonian
which is required by various numerical approaches --  such as
Discretized Light-Cone Quantization.

Of more topical interest is the application 
of these ideas to gauge theories in $D \geq 3$ 
space-time dimensions. It has been known for
some time that light-cone quantized gauge theories
possess particularly simple ladder relations 
if we work in the light-cone gauge $A_- = 0$ \cite{coll,abd,pin,
dall2,houch}.
The recursive procedure that enabled us to eliminate 
higher Fock sector dependence by introducing effective interactions
is also applicable in the context of these gauge theories.
This is the subject of current investigations.

Finally, we  remark that the 
Tamm-Dancoff truncation procedure is inherently
perturbative, and so intrinsically non-perturbative 
properties such as chiral symmetry breaking
and confinement are unlikely to emerge in such
an approximation scheme. In particular, in 
a numerical rendering of the 
bound state
equations, particle truncation is absolutely
necessary. Therefore,   
more analytical
work on the  {\em non-perturbative} properties of the bound 
state equations will be necessary to `enhance'
current numerical approaches\footnote{
The first steps toward 
enhancing the Discrete Light-Cone Quantization technique
appears in \cite{bret}.}. In the 
continuum formulation advocated here, 
achieving this is tantamount to summing all order contributions.
Whether this is a feasible task remains to be seen.

\begin{large}
{\bf Acknowledgements:}
\end{large}
 The author is grateful for many
fruitful interactions with S.Dalley and S.Brodsky
on the topic of small-$x$ relations, and
for the hospitality of H.-C.Pauli and
H.-J.Pirner. I have also benefited
from many stimulating discussions with M.Burkardt, 
E. Gubankova, S.Pinsky,
D.Robertson and S.Tsujimaru.

\vfil

\end{document}